\documentclass[]{elsart}
\usepackage{graphics}
\usepackage{graphicx}
\usepackage{epsfig}
\usepackage{amssymb}
\begin{document}
\begin{frontmatter}
\title{The mechanism of double exponential growth in hyper-inflation}


\author{Takayuki Mizuno\thanksref{fff}$^{a}$},
\author{Misako Takayasu$^{b}$},
\author{Hideki Takayasu$^{c}$}

\address{$^{a}$Department of Physics, Faculty of Science and Engineering, Chuo University, Kasuga, Bunkyo-ku, Tokyo 112-8551, Japan}
\address{$^{b}$Department of Complex Systems, Future University-Hakodate, 116-2 Kameda-Nakano-cho, Hakodate, Hokkaido 041-8655, Japan}
\address{$^{c}$Sony Computer Science Laboratories Inc., 3-14-13 Higashigotanda, Shinagawa-ku, Tokyo 141-0022, Japan}

\thanks[fff]{Corresponding author.\\
{\it E-mail address:}\/ mizuno@phys.chuo-u.ac.jp (T.Mizuno)}

\begin{abstract}
Analyzing historical data of price indices we find an extraordinary growth phenomenon in several examples of hyper-inflation in which price changes are approximated nicely by double-exponential functions of time. In order to explain such behavior we introduce the general coarse-graining technique in physics, the Monte Carlo renormalization group method, to the price dynamics. Starting from a microscopic stochastic equation describing dealers' actions in open markets we obtain a macroscopic noiseless equation of price consistent with the observation. The effect of auto-catalytic shortening of characteristic time caused by mob psychology is shown to be responsible for the double-exponential behavior.
\end{abstract}


\end{frontmatter}

\section{Introduction}
In ordinary markets prices fluctuate up and down fairly randomly as typically observed in foreign exchange markets or in stock markets. Inflation is the special economic situation in which prices apparently move monotonically upward and the value of money decreases rapidly. There are many examples of inflation not only in the history [1] but even now some countries are facing with the fear of inflation. Especially so-called the hyper-inflation is a kind of breakdown of currency system devastating the whole society. Detection of hyper-inflation in its early stage might contribute to avoid the tragedy, however, no such tool exists at present as scientific description of inflation has not been established yet.

\section{Empirical laws of inflation}
The worst inflation in the history occurred in Hungary right after the world war II. The exchange rate of 1 US dollar was about 100 Pengo in the beginning of July 1945. After one year the exchange rate became $6 \times 10^{24}$ Pengo in the middle of July 1946. Here, the inflation rate estimated by exchange rate difference is of order of $10^{22}$ per year. This terrible inflation was stopped by the introduction of the present Hungarian currency Forint in July 1946 which was exchangeable with Pengo with the exchange rate 1 Forint = $4 \times 10^{29}$ Pengo, which exceeds the symbol of large numbers, the Avogadro number, $6 \times 10^{23}$ !.

The time evolution of exchange rate of US dollar during this period is shown in Fig.1 in semi-log scale. As known from this figure the exchange rate grew exponentially until $t=220$ days measuring from 1st July 1945, and after that the growth became obviously faster than an exponential function. At that time, government of Hungary introduced a new unit of currency called the Adopengo. During the hyperinflation these two currencies, Pengo and Adopengo, coexisted in the market. The utility value of the Pengo was decreased by appearance of the new currency. We guess that the state of two currencies enhanced the decline of Pengo. In Fig.2 the exchange rates for $t>220$ are re-plotted in double logarithmic scale. The points are clearly on a straight line, demonstrating that the rapid growth is nicely approximated by a double exponential function.
\begin{equation}
p(t) \propto e^{a_1 t},\ \ \ \ (t < 220)
\end{equation}
\begin{equation}
\hspace{12mm} e^{b_1 e^{b_2 t}},\ \ (t > 220)
\end{equation}
where $a_1$, $b_1$ and $b_2$ are positive constants.

The single exponential growth is a generic property of inflation and it can be found widely in the historical data. The double exponential growths are rare events, however, we found 6 other examples as listed in Table 1. In economics the terminology "hyper-inflation" is used in rather rough sense to specify very heavy inflation. An example of definition is "inflation rates per month exceed 50\%" [2]. All examples of the double exponential growth in Table 1 belong to the worst extremes in the hyper-inflation category.

\begin{figure}
\begin{center}
\resizebox{8.525cm}{5.5cm}{\includegraphics{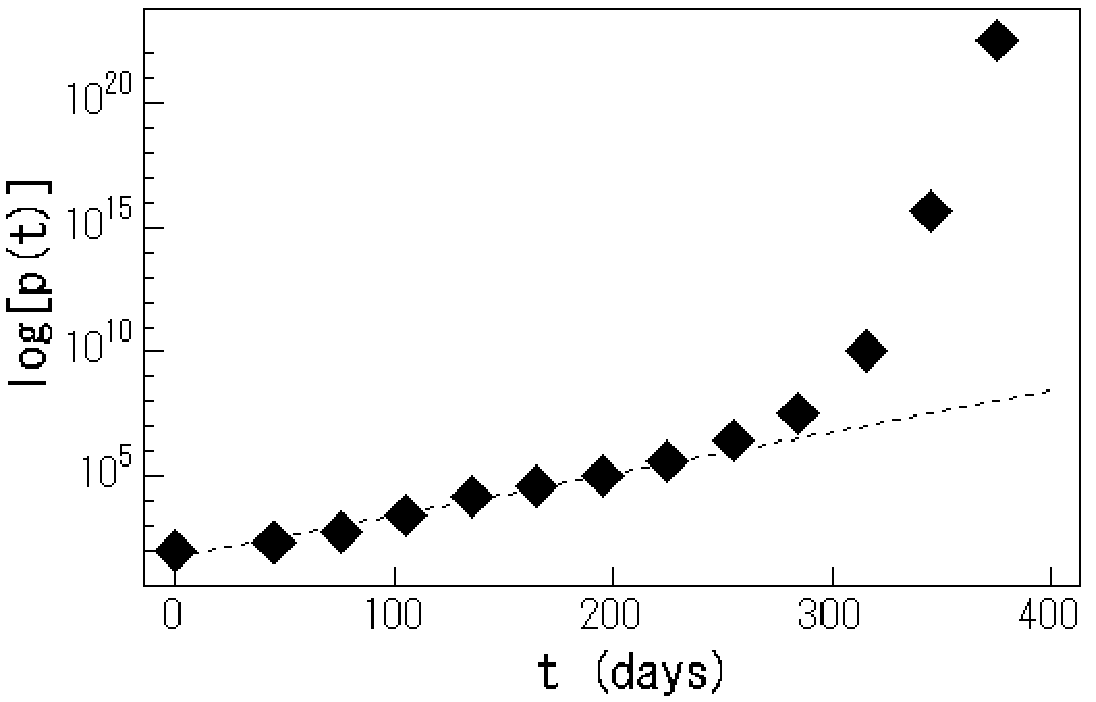}}
\end{center}
\caption{Semi-log plot of exchange rates of Pengo for 1 US dollar. The origin of the time axis is July 1st 1945. The straight line shows an exponential growth fitted for $t<220$.}
\label{fig1}
\end{figure}

\begin{figure}
\begin{center}
\resizebox{8.525cm}{5.5cm}{\includegraphics{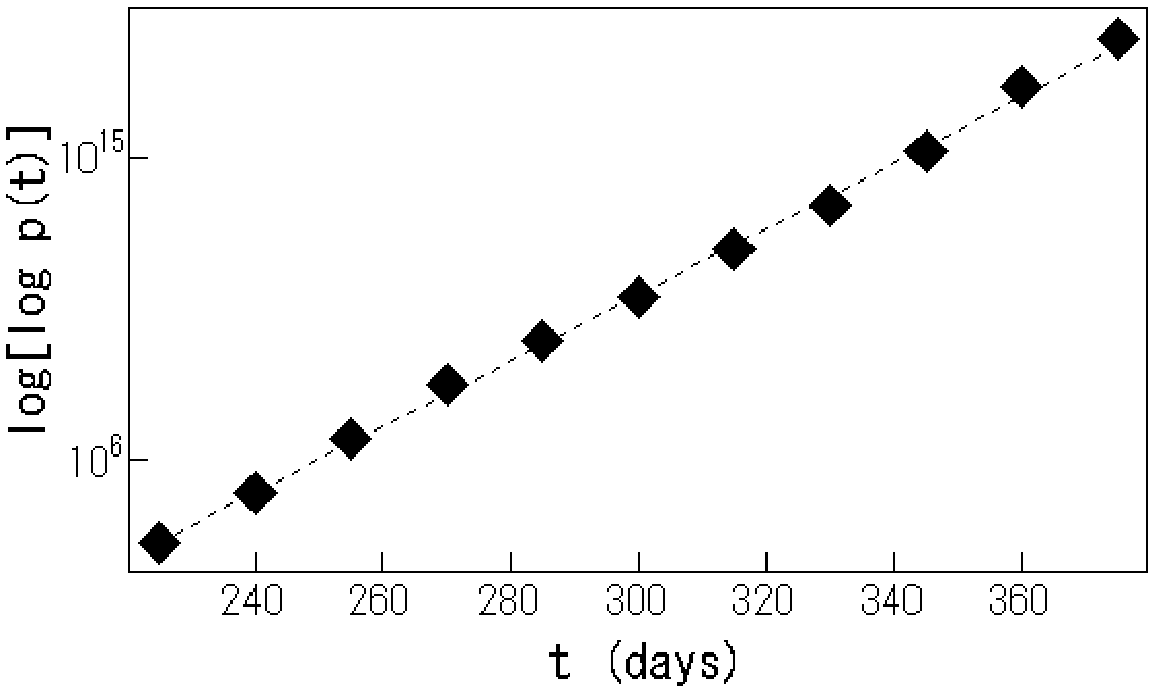}}
\end{center}
\caption{The part of $t>220$ of Fig.1 is re-plotted in double logarithmic scale. The straight line shows a double exponential fitting.}
\label{fig2}
\end{figure}

\begin{figure}
\begin{center}
\resizebox{8.525cm}{5.5cm}{\includegraphics{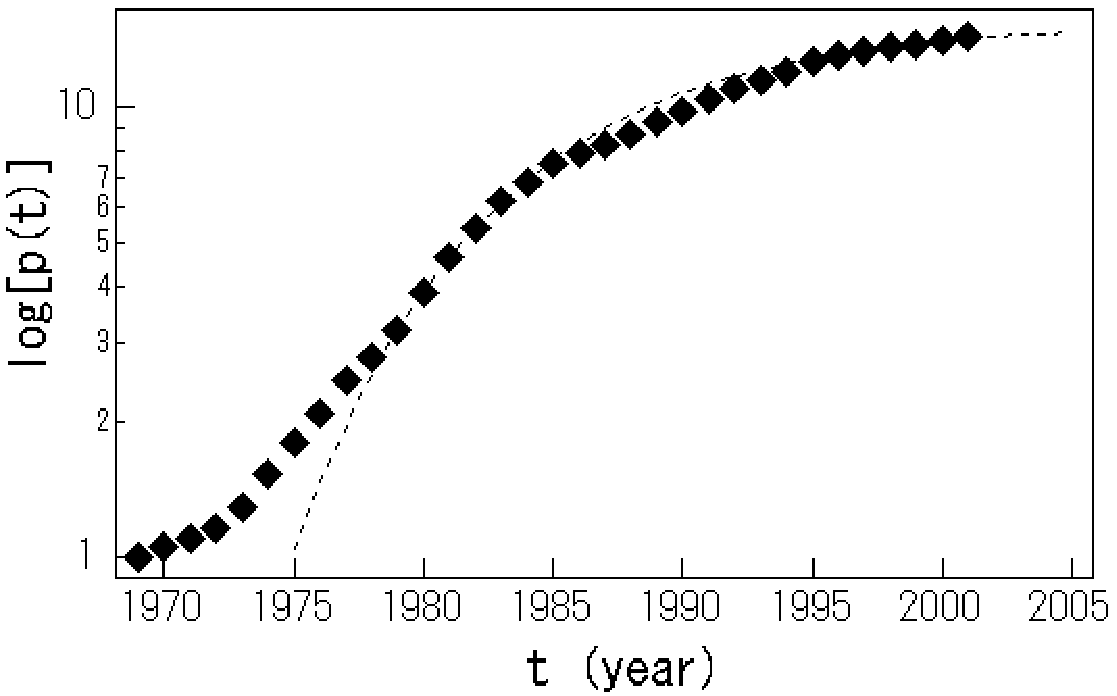}}
\end{center}
\caption{Semi-log plot of price index in Italy. The dotted line shows the theoretical curve fitted by a negative double logarithmic function, Eq.(3).}
\label{fig3}
\end{figure}

\begin{table} \caption{Examples of double exponential growth. The value of $B$ denotes the expoenent $B$ in Eq.(7). The value of $b_2$ in Eq.2 is given by $b_2=\ln(B^{1/2})$.}
\label{table1}
\begin{center}
\begin{tabular}{llcc}
\textbf{Country}& \textbf{Currency}& \textbf{Period}& \textbf{$B$($ \Delta t=1$year)}\\
BOLIVIA    & bolivianos & 1970 - 1985           & 2.8 \\
PERU       & intis      & 1970 - 1990           & 1.4 \\
ISRAEL     & sheqalim   & 1970 - 1985           & 1.4 \\
BRAZIL     & cruzeiro   & 1970 - 1994           & 1.3 \\
NICARAGUA  & cordoba	   & 1985 - 1991           & 1.3 \\
HUNGARY    & pengo      & 1946 Jan. - 1946 Jul. & $2.7^*$ \\
GERMANY    & mark       & 1922 Nov. - 1923 Oct. & $2.4^*$
\end{tabular}

*: The value of $B$ for $ \Delta t=1$ month (the value for 1 year is estimated by $B^{12}$).

\end{center}

\end{table}

\begin{table}
\caption{Examples of negative double exponential growth. The value of $B$ denotes the expoenent $B$ in Eq.(7). The value of $c_2$ in Eq.3 is given by $c_2=-\ln(B^{1/2})$.}
\label{table2}
\begin{center}
\begin{tabular}{llcc}
\textbf{Country}& \textbf{Currency}& \textbf{Period}& \textbf{$B$($ \Delta t=1$year)}\\
ITALY         & lire    & 1980 - 2001 & 0.8 \\
SOUTH AFRICA  & rand    & 1986 - 2000 & 0.8 \\
EGYPT         & pound   & 1992 - 2000 & 0.7 \\
PORTUGAL      & escudos & 1984 - 1997 & 0.7 \\
ICELAND       & kronur  & 1983 - 1995 & 0.6 \\
GUINEA        & francs  & 1980 - 1997 & 0.6
\end{tabular}
\end{center}

\end{table}

Another functional form of inflation growth can be found at the ending of inflation period. In the case of Hungary the inflation stopped quite suddenly, however, there are cases that inflation stabilized rather slowly. In Fig.3 the growth of annual price index in Italy is plotted in semi-log scale. Here, the local slope of the price index becomes gentler year by year. The data points from 1978 to 2000 are well approximated by the following negative double exponential function as fitted by the dotted line:
\begin{equation}
p(t) \propto e^{-c_1 e^{-c_2 t}}  ,
\end{equation}
where $c_1$ and $c_2$ are positive constants. There are many examples of this type of relaxations as listed in Table 2.

It is known that the inflation phenomenon is mainly caused by widespread of so-called "inflation mind" which is the expectation of general people that prices will rise in the near future [3]. The central bank's excess supply of money can contribute to inflation, however, hyper-inflation can not be realized without general people's inflation mind because the effect of cash supply is limited in the whole economy. In fact, in the case of Austria (1921-1924), the price index stabilized automatically although the central bank was keeping an excess money supply policy [4].

\section{Stochastic dynamics of market prices}
In 1956 Cagan[2] introduced a set of evolution equations of prices for description of inflation, which are consisted of market price $p(t)$ and the people's averaged expectation price $p^*(t)$. In the theory these two prices evolve due to the linear positive feedback mechanism; an upward change of market price in a unit time induces rise in the people's expectation price, and such anticipation pulls up the market price EEE. As a result the exponential growth of market price in inflation can be naturally explained, however, the theory fails to explain the double exponential behaviors. In order to explain the hyper-inflation, Cagan assumed that the inflation rate is dependent on the rate of increase of money supply, and introduced a model of hyper-inflation. However, as mentioned above the effect of money supply is known to be limited, so in the following discussion we neglect the effect of governmental policy of money supply, instead we introduce an effect of open market fluctuations.

We introduce the following set of stochastic equations of $p(t)$ and $p^*(t)$ as a generalization of Cagan's approach:
\begin{equation}
p(t+ \Delta t)/p(t)=(p^*(t)/p(t))^{A(t)}e^{f(t)}
\end{equation}
\begin{equation}
p^*(t+ \Delta t)/p^*(t)=(p(t)/p(t- \Delta t))^{B(t)}e^{f^*(t)} 
\end{equation}
Here, the last terms, $f(t)$ and $f^*(t)$, represent random noises, and the original Cagan's equation is obtained for the special case that $A(t)=1$ and $B(t)=1$ with no noises. The meaning of the coefficients $A(t)$ and $B(t)$ can be explained as follows. By taking logarithm and expanding the variables assuming that  $\Delta t$ is close to 0, Eqs.(4) and (5) coincide to the known set of linear stochastic equations describing microscopic properties of market price changes [5].

The coefficient $A(t)$ is equivalent to the inverse of price rigidity characterizing the market price response to the change of demand and supply. The value of $A(t)$ is larger than 1 when market orders are rich, such as the case that many people rush in the market. The coefficient $B(t)$ is called as the dealer's response to the market price changes which characterizes people's averaged expectation of the future price. The value of $B(t)$ is larger than 1 if people expects larger price change in the near future than that of present market price change. Various market behaviors can be characterized by these coefficients, for example, markets are stable when both $A(t)$ and $B(t)$ are less than 1, bubbles and crashes occur when $A(t)<1$ and $B(t)>1$, and damping oscillation can be found when $A(t)>1$ and $B(t)<1$ [6]. The widely observed open market property that price changes follow a distribution with fat-tails close to a power law [7] is known to be attributed to the random fluctuation of the multiplicative coefficient $B(t)$ [8]. An approach of real time characterization of market condition in terms of $A(t)$ and $B(t)$ is also in progress [9].

\section{Derivation of macroscopic market equation using renormalization}
In the bubble behaviors prices grow exponentially with time, which are similar to the observed exponential growth in the inflation data. This implies that the bubble behaviors in markets may have a similar mathematical structure with the inflation although the time scales can be very different. In order to derive a macro-scale dynamics of prices, we apply the idea of renormalization group method developed in statistical physics to the price equations.

The Monte Carlo renormalization group method is based on coarse-graining procedures as follows: First, we fix the parameters such as the mean values and variances of the random variables ${{A(t), B(t), f(t), f^*(t)}}$ in Eqs.(4) and (5) with a small time step $\Delta t$. We numerically simulate price changes repeatedly by using different random numbers. Then, we observe the resulting price changes in a discrete manner with a large time resolution $\Delta t'$. Assuming that the large scale price changes follow the same form of equation with different combination of parameters, we estimate the map from the parameter sets for small $\Delta t$ to those for large $\Delta t'$. Using this map the parameter values for the macroscopic dynamics can be deduced by repeating the mapping in the parameter space.

\begin{figure}
\begin{center}
\resizebox{8.525cm}{14cm}{\includegraphics{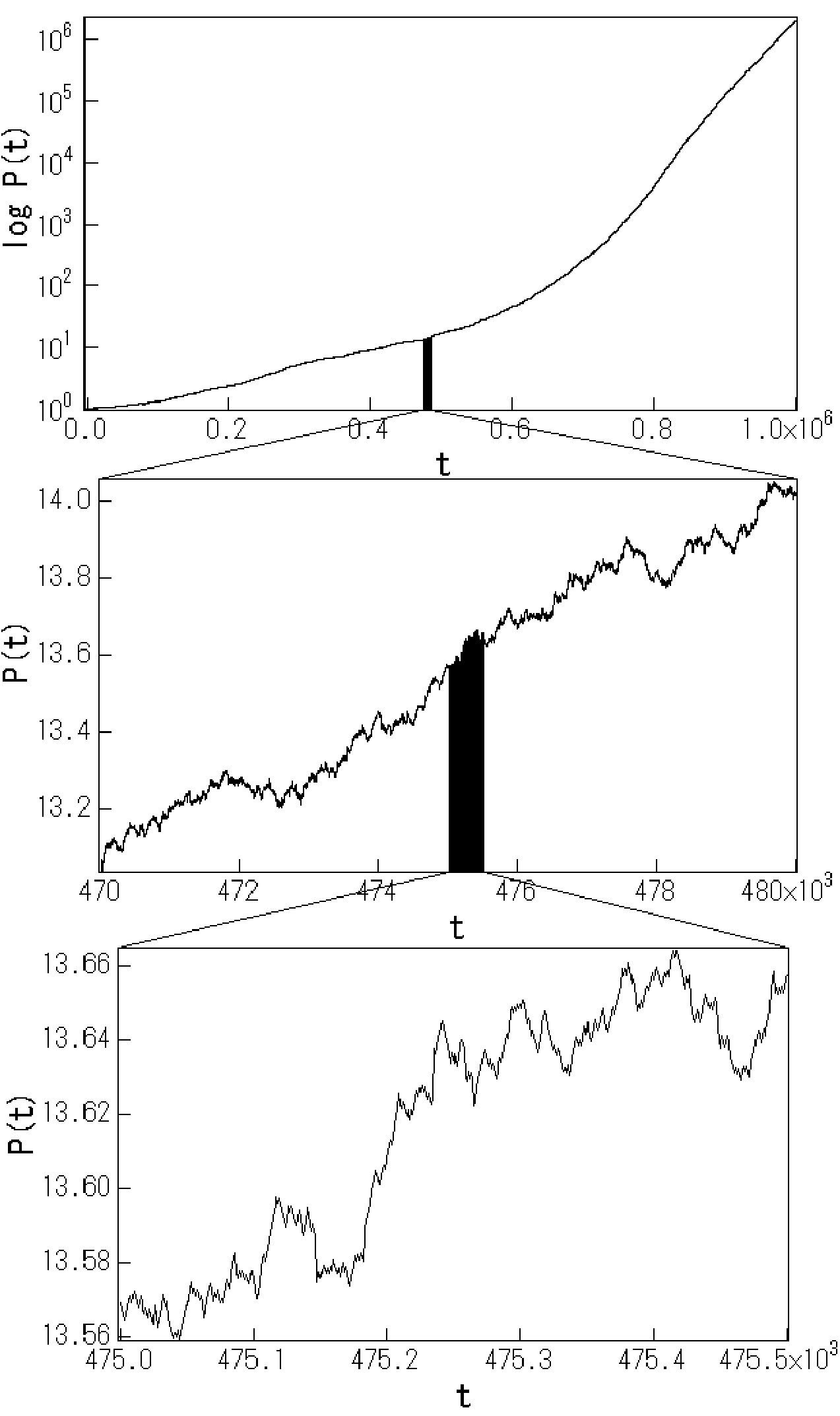}}
\end{center}
\caption{An example of coarse-grained view of price fluctuations. Numerical simulation is done in the microscopic scale as shown in the bottom. Observing the price fluctuations in a larger scale (the middle), the fluctuations become relatively smaller, and in the macroscopic level (the top) the price motion looks following a smooth dynamics. Note that microscopic price fluctuates up and down even in a hyper-inflation state.}
\label{fig4}
\end{figure}

By repeating the mapping we can find that the price dynamics converges to the following simple combination of non-random parameter values for a wide range of initial parameters in the large-scale limit:
\begin{equation}
A(t)=1, B(t)=B, f(t)=0, f^*(t)=0        ,
\end{equation}
where $B$ is a non-trivial positive constant. This result means that the random noise terms in Eqs.(4) and (5) are negligible in the large scale limit as intuitively demonstrated in Fig.4.

 From Eqs.(4) and (5) with the parameter values given by Eq.(6) we have the following simple deterministic nonlinear equation for the macroscopic price evolution:
\begin{equation}
p(t+2 \Delta t)/p(t+ \Delta t)=(p(t)/p(t- \Delta t))^B .
\end{equation}
This equation is solved as the following,
\begin{equation}
p(t) \propto \left\{
\begin{array}{ll}
e^{a_1 t}, & B=1 \\
e^{b_1 e^{b_2 t}}, & B>1 \\
e^{-c_1 e^{-c_2 t}}, & B<1
\end{array}
\right.
\end{equation}
where 
$a_1={\Delta t}^{-1} \log {\frac{P(\Delta t)}{P(0)}}$,
$b_1= \frac {B^{\frac{1}{2 \Delta t}}}{B^{dt}-1} \log \frac {P( \Delta t)}{P(0)}$, $b_2=\log B$,
$c_1= -\frac {B^{\frac{1}{2 \Delta t}}}{B^{dt}-1} \log \frac {P( \Delta t)}{P(0)}$, $c_2=-\log B$
.
We have the exponential price growth consistent with Eq.(1) when $B=1$, while for $B>1$ the solution becomes the double exponential function of time identical to Eq.(2). In the case of $B<1$ gives the price evolution of the negative double exponential function of Eq.(3) that converges to a stable price level. These three types of price motions are the basic behaviors in the large-scale limit consistent with the observation in the historical data.

\section{Discussion}
The meaning of Eq.(7) can be recognized more clearly by introducing a new quantity T(t) which is the time interval needed to make the price double:
\begin{equation}
p(t+T(t)) = 2p(t)          .
\end{equation}
Considering a continuum limit we can show that $T(t)$ satisfies the following differential equation:
\begin{equation}
\frac {dT(t)}{dt} \propto (1-B)T(t).
\end{equation}
Obviously, $T(t)$ is constant for $B=1$ and it decays or grows exponentially with time for $B>1$ and $B<1$, respectively. Such functional dependence is directly checked from the inflation data of Hungary as shown in Fig.5. We can confirm an exponential shrink of $T(t)$ for the double exponential period $t>220$ as expected.

This quantity, $T(t)$, characterizes the speed of time in the mob psychology in the following meaning. In the situation of an ordinary exponential inflation people's clock speed is nearly constant. When the price begins to grow faster than the expected inflation rate, then the people's clock speed is accelerated auto-catalytically and the society falls into the double exponential hyper-inflation phase.

 By generalizing the renormalization technique introduced in this paper it may be possible to establish the method of estimating the key parameter $B$ in Eq.(7) from given data of microscopic market price fluctuations. Such a technique will hopefully pave the way to conquer the fear of inflation in a scientific manner.
\begin{figure}
\begin{center}
\resizebox{8.525cm}{5.5cm}{\includegraphics{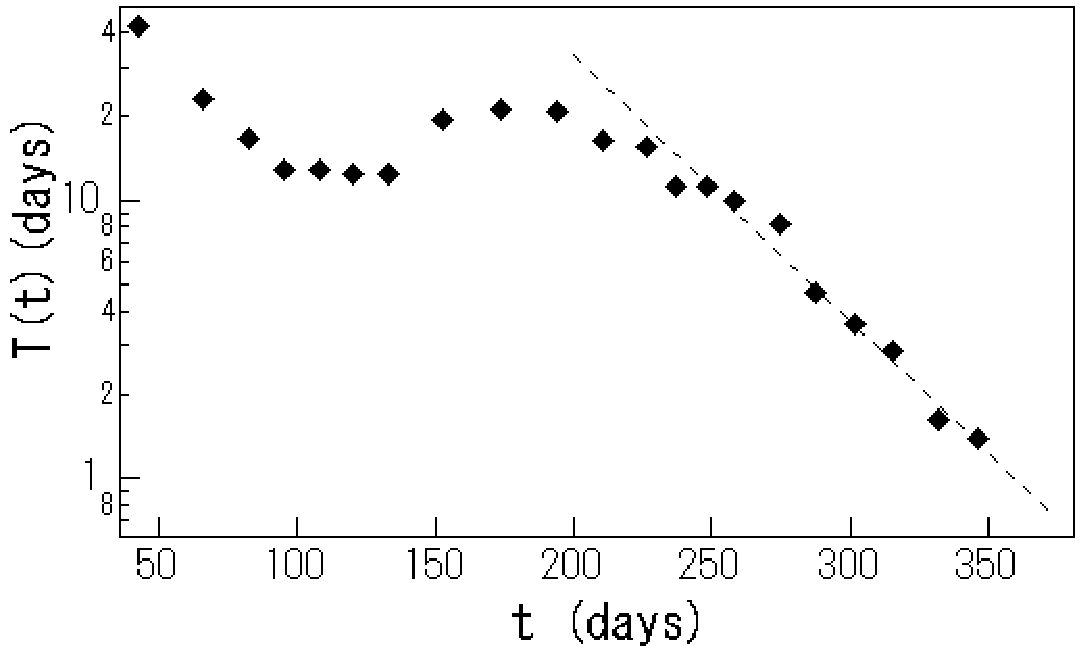}}
\end{center}
\caption{The change of time interval (days) needed to double the price in Hungary 1946. The abscissa is the same as in Fig.1. The straight line shows an exponential decay consistent with Eq.(10).}
\label{fig5}
\end{figure}

\vspace{1cm}
\textbf{Acknowledgments}

We would like to appreciate Dr. M. Vicsek for providing us some historical data of Hungarian currency, Dr. M. Katori and S.Kurihara for stimulus discussions.





\begin{thebibliography}{}


\bibitem{1}
The World Economic Outlook (WEO) Database September 2000.\\
(http://www.imf.org/external/pubs/ft/weo/2000/02/data/index.htm)

\bibitem{2}
P.\ Cagan, The Monetary Dynamics of Hyperinflation. In Milton Friedman (ed.), {\it Studies in the Quantity Theory of Money} (University of Chicago Press, 1956).
\bibitem{3}
T.\ Sargent, {\it Rational Expectations and Inflation} (Harper and Row Publishers, New York, 1986).

\bibitem{4}
H.\ Huziki, {\em Monetary and Economic Studies} 19 2 (June, 2000) 31-72, Bank of Japan

\bibitem{5}
H.\ Takayasu and M.\ Takayasu, {\em Physica A} 269 (1999) 24.

\bibitem{6}
H.\ Takayasu and M.\ Takayasu, {\em Econophysics - Toward Scientific Reconstruction of Economy} (in Japanese, Nikkei, Tokyo, 2001). 

\bibitem{7}
R.\ N.\ Mantegna and H.\ E.\ Stanley, {\it An Introduction to Econophysics: Correlation and Complexity in Finance} (Cambridge University Press, Cambridge, 2000).

\bibitem{8}
H.\ Takayasu, A.-H.\ Sato and M.\ Takayasu, {\em Phys. Rev. Lett.} 79 (1997) 966.

\bibitem{9}
M.\ G.Li, A.\ Oba and H.\ Takayasu, in {\em Empirical Science of Financial Fluctuation - The Advent of Econophysics}, (Springer Verlag, Tokyo, 2002), 260-270. 


\end{thebibliography}
\end{document}